\documentclass[preprintnumbers, floatfix, letterpaper, twocolumn,aps,prd,epsfig,nofootinbib,natbib,longbibliography]{revtex4-1}
\usepackage{graphicx}
\usepackage{epstopdf}
\usepackage{latexsym}
\usepackage{amssymb}
\usepackage{amsmath}
\usepackage{color}
\usepackage{mathrsfs}
\usepackage{xparse}
\usepackage{bbding}
\usepackage{pifont}
\usepackage{comment}
\usepackage{ulem}
\usepackage{float}
\usepackage[inline]{enumitem}
\usepackage{soul}
\usepackage[caption=false]{subfig}

\let\Right\right
\let\Left\left
\makeatletter
\def\right#1{\Right#1\@ifnextchar){\!\right}{}}
\def\left#1{\Left#1\@ifnextchar({\!\left}{}}
\makeatother

\usepackage[
            pdfstartview=FitH,
            bookmarksnumbered=true,
            bookmarksopen=true,
            colorlinks,
            linkcolor=blue,
            anchorcolor=green,
            citecolor=blue
            ]{hyperref}
\begin{document}

  \renewcommand\arraystretch{2}
 \newcommand{\bq}{\begin{equation}}
 \newcommand{\eq}{\end{equation}}
 \newcommand{\bqn}{\begin{eqnarray}}
 \newcommand{\eqn}{\end{eqnarray}}
 \newcommand{\nb}{\nonumber}
 \newcommand{\lb}{\label}
 
\newcommand{\La}{\Lambda}
\newcommand{\va}{\scriptscriptstyle}
\newcommand{\be}{\nopagebreak[3]\begin{equation}}
\newcommand{\ee}{\end{equation}}

\newcommand{\ba}{\nopagebreak[3]\begin{eqnarray}}
\newcommand{\ea}{\end{eqnarray}}

\newcommand{\la}{\label}
\newcommand{\n}{\nonumber}

\newcommand{\R}{\mathbb{R}}

 \newcommand{\cb}{\color{blue}}
    \newcommand{\cc}{\color{cyan}}
        \newcommand{\cm}{\color{magenta}}
\newcommand{\rc}{\rho^{\scriptscriptstyle{\mathrm{I}}}_c}
\newcommand{\rd}{\rho^{\scriptscriptstyle{\mathrm{II}}}_c} 
\NewDocumentCommand{\evalat}{sO{\big}mm}{%
  \IfBooleanTF{#1}
   {\mleft. #3 \mright|_{#4}}
   {#3#2|_{#4}}%
}
\newcommand{\PRL}{Phys. Rev. Lett.}
\newcommand{\PL}{Phys. Lett.}
\newcommand{\PR}{Phys. Rev.}
\newcommand{\CQG}{Class. Quantum Grav.}

\title{Quasi-Normal Modes of Loop Quantum Black Holes Formed from Gravitational Collapse}

\author{Chao Zhang${}^{a, b}$}
\email{ phyzc@cup.edu.cn}

\author{Anzhong Wang${}^{c}$}
\email{ anzhong$\_$wang@baylor.edu; Corresponding author}

\affiliation{${}^{a}$Basic Research Center for Energy Interdisciplinary, College of Science, China University of Petroleum-Beijing, Beijing 102249, China\\
${}^{b}$Beijing Key Laboratory of Optical Detection Technology for Oil and Gas, China University of Petroleum-Beijing, Beijing 102249, China\\
${}^{c}$ GCAP-CASPER, Physics Department, Baylor University, Waco, Texas 76798-7316, USA}

\date{\today}

\begin{abstract}

In this paper, we study the quasi-normal modes (QNMs) of a scalar field in the background of a large class of quantum black holes that can be formed from gravitational collapse of a dust fluid in the framework of effective loop quantum gravity. The loop quantum black holes (LQBHs) are characterized by three free parameters, one of which is the mass parameter, while the other two are purely due to quantum geometric effects. Among these two quantum parameters, one is completely fixed by black hole thermodynamics and its effects are negligible for macroscopic black holes, while the second parameter is completely free (in principle). In the studies of the QNMs of such LQBHs, we pay particular attention to the difference of the QNMs between LQBHs and classical ones, so that they can be observed for the current and forthcoming gravitational wave observations, whereby place the LQBH theory directly under the test of observations.

\end{abstract}


\maketitle
\section{Introduction}

\renewcommand{\theequation}{1.\arabic{equation}} \setcounter{equation}{0}

Black holes (BHs) are one of the most mysterious phenomena in the universe. The existence of BHs provides us a perfect way to test gravitational effects under extremely
 strong gravitational fields, such as the formation of gigantic jets and disruption of neighboring stars. On the other hand, from the theoretical point of view, BHs
are also excellent labs to test theories of gravity that are
different from general relativity (GR) (see, e.g., \cite{test_GR1,test_GR2,test_GR3,Xiang2019,Chao2020,Chao2020b,Chao2023a,Zack2020}). It is true that BHs were initially only discovered as mathematical solutions of GR \cite{Schwarzschild1916}. Nonetheless, observations have already confirmed their existence, including those from gravitational wave (GW) observations \cite{Ref1}, which marks the beginning of a new era — the GW astronomy.

Following the first observation of GW, more than 90 GW events, which are all from coalescences of compact bodies, black holes and/or neutron stars, have been identified by the LIGO/Virgo/KAGRA (LVK) scientific collaborations  \cite{GWs, GWs19a, GWs19b, GWsO3b}. In the future, more advanced ground- and space-based GW detectors will be constructed \cite{Moore2015, Gong:2021gvw}, such as Cosmic Explorer \cite{CE}, Einstein Telescope \cite{ET}, LISA \cite{LISA}, TianQin \cite{Liu2020, Shi2019}, Taiji \cite{Taiji2}, and DECIGO \cite{DECIGO}. These detectors will enable us to probe signals with a much wider frequency band and long distances.  This has triggered the interest in the observation of quasi-normal mode (QNM) of black holes \cite{Berti2009, Berti18}, extreme mass ratio inspirals (EMRIs) \cite{Gong2023, Tu2023}, the effect of dark matter on GWs \cite{Chao2022,Shen2024}, to name only a few of them.

GWs emitted during the ringdown stage of a coalescence,  where a massive black hole is just formed, can be considered as the linear combination of the QNMs of the just formed BH \footnote{The BH just formed from the merge stage of two compact objects is highly nonlinear, and the linear perturbation theory cannot be applied to the initial stage of such just formed BH. Thus, how to connect the ringdown waveform determined by QNMs is still an open question \cite{Cheung2023, Mitman2023}. In this paper, we shall not consider such a matching, and simply consider only the times during which the GWs can be well expressed as linear combinations of QNMs.}, and are in general studied with the
perturbation theory \cite{Regge1957, Teukolsky1973, Will1985, Leaver1985, Iyer1987,Chandra92}. QNMs of such perturbations,  including scalar, vector (electromagnetic), and tensor (gravitational), have been extensively studied in GR as well as in various modified theories of gravity. From the theoretical point of view, QNMs are eigenmodes of
dissipative systems and usually contain two parts, the real part and the imaginary part. Its real part gives the frequency of vibration while its imaginary part provides the damping time. The information contained in QNMs provide the keys in revealing whether BHs are ubiquitous in our universe, and more important whether GR is the correct theory to describe gravity even in the strong field regime. As a matter of fact, according to the no-hair  theorem of GR, BHs are described uniquely by only three parameters, mass, charge and angular momentum.
Thus, in GR QNMs are functions of only these three parameters. Hence, observing such modes can directly test such relations. For more details, we refer readers to \cite{Berti2009, Berti18, Kokkotas1999, Konoplya2011}.

On the other hand, from the observational point of view, the sensitivities of the current detectors limit the detection of QNMs mainly to the dominant modes, and it is still an ongoing debate on whether other modes can be observed by LIGO, Virgo and KAGRA \cite{Carullo2018,Isi2019,Bustillo2021,Finch2022,Ma2023,Isi2023, Carullo2023,Capano2023}. However, with the technological advances, it is strongly believed that the upcoming third-generation detectors, either the ground-based or the space-based ones, will be able to detect various sub-leading modes, see for example, \cite{Shi2024} and references therein.

In this paper, we shall focus on the QNMs of a large
class of quantum BHs in the context of effective loop
quantum gravity (LQG) \cite{Lewandowski2023,Luca2024}. It should be noted that QNMs of loop quantum black holes (LQBHs) \footnote{Loop quantum black holes have been studied extensively in the past decades or so. For details, see \cite{Gambini2023,Ashtekar2023,OG2024,GZO2024} and references therein.} have been already studied extensively \cite{Tu2023,Yan2023,Yan2022,Liu2021,Ramin2021,Arbey2021,Livine2024,Xia2024,Fu2023,Shao2024,Yang2023,Okounkova2020,LiuC2020,Cao:2024oud}. However, in most of
these models, the observational effects are negligible for
macroscopic BHs. The LQBHs to be studied in this paper
are distinguishable from the previous ones in the senses:
(a) They can be directly formed from the gravitational
collapse of a spherically symmetric dust cloud; and (b)
the solutions usually contain three free parameters, one
is the mass of the BH, and the other two characterize
the quantum geometric effects. Among the two quantum
parameters, one is completely fixed by the black hole
thermodynamics, whose observational effects are always
negligible for macroscopic BHs. On the other hand, the
second quantum parameter is free \cite{Luca2024}. Our major
goal here is to study its effects to QNMs, in order to
obtain observational constraints on this second quantum
parameter.

The rest of the paper is organized as follows: Sec.\ref{secII} describes
 briefly the LQBH solutions we are going to study
in this paper and the equations for a scalar field moving
 in such backgrounds. The latter leads to the master
equation for the study of QNMs. Knowing this, in Sec.\ref{secQNM} 
we calculate the QNMs for both the time- and frequency-domains
 since the former reflects a comprehensive effect
of various different modes (to be introduced below), while
the latter provides a more quantitative evidence for the
confinement to the coupling-parameter phase space with
future GW missions. Some of our concluding remarks
and future outlooks are given in Sec.\ref{secconclusion}.

Before proceeding further, we would like to note
that linear perturbations of spherically symmetric LQBH
spacetimes have not been worked out yet in the framework
 of LQG, although some important steps were already
 taken \cite{Mena2024}. In particular, the QNMs of the scalar
perturbations studied in this paper cannot be compared
directly with the QNMs of spin-2 gravitons of black
holes studied in GR, which are directly related to the
gravitational waveforms emitted in the ringdown phase
\cite{Berti2009,Berti18,Kokkotas1999,Konoplya2011}. So, in this paper, what we are comparing
with are the QNMs of a scalar field moving in the backgrounds
 of LQBHs and the QNMs of the same scalar field
moving in the Schwarzschild BH background of GR, as
we believe that the deviations of the scalar perturbations
between these two theories are of the same order as that
of the deviations of the spin-2 gravitons between the two
same theories.

In addition, in this paper we are adopting the unit system so that $c=G=\hbar =1$, where $c$ denotes the speed of light, $G$ is the gravitational constant, and $\hbar$ the Planck constant divided by $2\pi$.
All the Greek letter in indices run from 0 to 3. The other usages of indices will be described at suitable places. 


\section{LQBH Background and scalar perturbation}
\renewcommand{\theequation}{2.\arabic{equation}} \setcounter{equation}{0}
\label{secII}

In this section, we shall first introduce briefly the background of a spherically symmetric loop quantum black hole formed from the gravitational collapse of a homogeneous and isotropic dust fluid in the framework of effective theory of loop quantum gravity\cite{Lewandowski2023,Luca2024}, and then consider the linear perturbations of a scalar field in such backgrounds \cite{Stashko2024}. In doing so, we ignore the back-reactions of the scalar field.

\subsection{Loop Quantum Black Holes formed from Gravitational Collapse}

The LQBH solution we are studying is given by \cite{Luca2024}
\bqn
\lb{ds2}
d s_{+}^2 &=& -f(r) d t^2+\frac{1}{f(r)} d r^2 + r^2 d {\Omega}^2,~~
\eqn
where
\bqn
\lb{fr}
f(r) &=& 1 - \frac{2 M}{r} + \frac{ \alpha}{r^2} \left(\frac{M}{r}+ \frac{B}{2} \right)^2,
\eqn
and $d \Omega^2 =  d \theta^2 + \sin^2{\theta} d \varphi^2$.
Here $B$ is a dimensionless coupling parameter.  Considering the above solution as describing the external spacetime of a collapsing ball,
while inside the ball the spacetime is described by the FLRW universe
\bqn
\lb{ds2B}
d s_{-}^2 &=& -d T^2+ a^2\left(dR^2  + \chi_k^2(R) d {\Omega}^2\right),~~
\eqn
it was found that the junction conditions lead to \cite{Luca2024}
\bqn
\lb{eqB}
B = - k \chi_k^2(R_0) = \begin{cases}
- \sin^2(R_0), & k = 1,\cr
0, & k = 0,\cr
\sinh^2(R_0), & k = -1,\cr
\end{cases}
\eqn
where $R_0$ denotes the junction surface. The factor $\alpha$, which has the dimension of $r^2$, is given by \cite{Chiou2008, Luca2024}
\bqn
\lb{alpha}
\alpha &=& 16 \sqrt{3} \pi \gamma^3 \ell_{\text{pl}}^2,
\eqn
where  $ \ell_{\text{pl}}$ denotes the Planck length, and $\gamma$ is known as the Barbero-Immirzi parameter whose value is set to $\gamma \approx 0.2375$ using black hole thermodynamics in LQG  \cite{Meissner2004}.

To calculate QNMs of the above BHs, let us first introduce the dimensionless coordinates $\tilde{t}$
and $\tilde{r}$
\bqn
\lb{eq2.4}
t = r_s \tilde{t}, \quad r = r_s \tilde{r},
\eqn
where $r_s \equiv 2M$. Then, in terms of $\tilde{t}$ and $\tilde{r}$, the metric reduces to
\bqn
\lb{eq2.5}
d s_{+}^2 &=& r_s^2 \left(- f(\tilde{r}) d\tilde{t}^2 + \frac{d^2\tilde{r}}{f(\tilde{r})} + \tilde{r}^2d^2\Omega\right)\nb\\
&\equiv& r_s^2 d\tilde{s}^2,
\eqn
where
\bqn
\lb{ftr}
f(\tilde{r}) &=& 1 - \frac{1}{\tilde{r}} + \frac{\tilde\alpha}{\tilde{r}^2} \left(\frac{1}{\tilde{r}}+ B\right)^2,\nb\\
\tilde{\alpha} &\equiv& \frac{\alpha}{4r_s^2} = \sqrt{3} \pi \gamma^3 \left(\frac{\ell_{\text{pl}}}{M}\right)^2.
\eqn
Note that for solar-mass BHs, we have $M \simeq 10^{38} \; \ell_{\text{pl}}$, so that (the dimensionless) 
$\tilde{\alpha} \simeq {\cal{O}}\left(10^{-77}\right)$. That is, for macroscopic BHs, the quantum gravitational effects from the $\alpha$ terms are negligible, unless the combination of $\tilde{\alpha} B$ is large, a condition that we will assume in order to compare the resultant QNMs of LQBHs with GW observations. Considering Eq.(\ref{eqB}), we find that this is possible only when $k = -1$.
As a result, the values of $B$ to be considered will be a non-negative number. 

 On the other hand, since $ds_+^2$ and $d\tilde{s}^2$ are related conformally with a constant conformal factor [cf., \eqref{eq2.5}], the physics of the spacetimes of $ds_+^2$ can be obtained directly from that of the spacetimes of $d\tilde{s}^2$.
In particular, by following the same manner of definition, we have $R = \tilde{R}/r_s^2$ and
\bqn
\lb{eq2.7}
\tilde \omega &\equiv& r_s \omega, 
\eqn
where $\omega$ and the dimensionless $\tilde\omega$ are the QNMs of $ds_+^2$ and $d\tilde{s}^2$, respectively, and $R$ and $\tilde{R}$ are their corresponding curvatures\footnote{One should avoid confusing the curvature $R$ in here with the $R$ coordinate appearing in \eqref{ds2B}.}. Therefore, in the rest of the paper we shall work with the spacetime described by $d\tilde{s}^2$, i.e. 
\bqn
\lb{eq2.8}
d\tilde{s}^2 &=& - f(\tilde{r}) d\tilde{t}^2 + \frac{d^2\tilde{r}}{f(\tilde{r})} + \tilde{r}^2d^2\Omega,
\eqn
when we calculate the QNMs of the quantum BHs. Then, using Eq.(\ref{eq2.7}) we can easily read off $\omega$ for a given BH with mass $M$.

 Moreover, by looking at Eq.\eqref{fr}, we immediately notice that a positive radius of the metric horizon $r_{MH}$, which is defined through $f(r=r_{MH})=0$, is absent at the $M \to 0$ limit. Therefore, by requiring $r_{MH}>0$, we have to look at the cases where $M$ is large enough. Keeping this in mind, in this paper we shall consider the cases of $\lambda \equiv M/M_{\odot}\in (3, 100)$\footnote{In order to study their observational results, in this paper we mainly consider BHs with masses in the order of the solar mass or larger.  In particular, the current GW observations set the lower limit to about $3 M_{\odot}$ (See., e.g., \cite{LVK2024}), which makes it reasonable for us to consider the range $M/M_{\odot}\in (3, 100)$. As will be seen later, the difference on QNMs between a LQBH and a Schwarzschild BH sensitively depends on the ratio $M/M_{\odot}$ when that is relatively small. After it has reached to about 100 (and even before that), the difference is observationallly negligible. That is why we set the upper limit to $100$ in here.} in this current paper, where $M_{\odot}$ denotes the solar mass.

\begin{figure}[h]
\includegraphics[width=0.9
\linewidth]{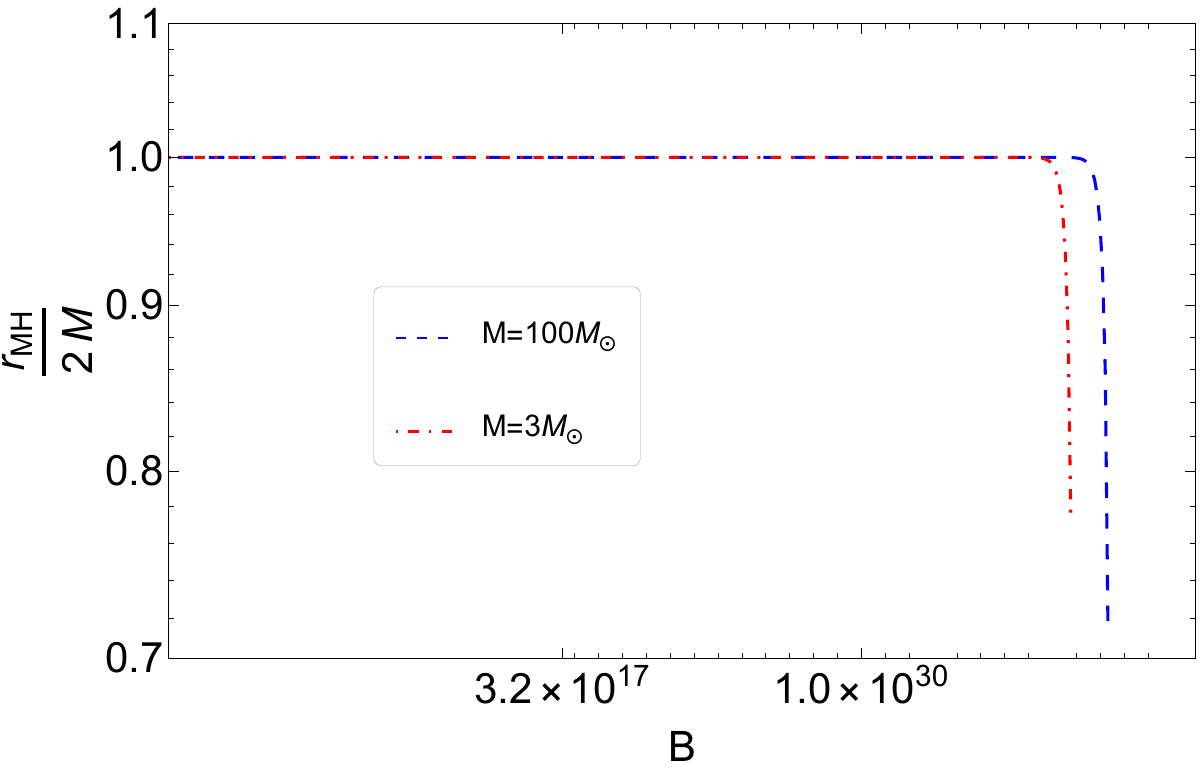} 
\caption{The behavior of the dimensionless quantity $r_{MH}/(2M)$ as a function of $B$ for different values of $M$.} 
	\label{plot1}
\end{figure}

Knowing this, we are on the position of determining an approximate upper limit for $B$ by requiring $r_{MH}>0$ for $\lambda \in (3, 100)$. 
By plotting out the behavior of $r_{MH}$ as a function of $B$ for the range of $M$ under consideration in Fig. \ref{plot1}, we can notice an approximate upper limit of $B$. A more careful study to the equation $f(r=r_{MH})=0$ shows that, for a positive $r_{MH}$ to exist for any $\lambda \in (3, 100)$, the upper limit of $B$ is about $B_{max} \approx 7.176 \times 10^{38}$. 
It is worth mentioning here that, to avoid confusions arising by using different unit systems, in Fig. \ref{plot1} we are focusing on the dimensionless quantity $r_{MH}/(2 M)$ (instead of the quantity $r_{MH}$ itself). Therefore, the curves of Fig. \ref{plot1} will not be changed when considering different unit systems. 

In addition, by observing Fig.\ref{plot1}, we also notice that the value of $r_{MH}/(2 M)$ will soon deviate from that of GR when $B$ reaches a critical point (For the $M/M_{\odot}=3$ case, that critical point is located around $B =10^{38}$, which is very close to the upper limit of $B$).  Before getting close to such a critical point, it is very hard to distinguish the value of $r_{MH}/(2 M)$ from its GR counterpart. This implies that one may find the observable difference between a LQBH and a Schwarzschild one only for sufficiently large $B$'s (which should be quite close to $B_{max}$). Yet, such a critical point is getting larger and larger as $M$ increases. Therefore, given an overall  $B_{max}$ (obtained from the $M/M_{\odot}=3$ case), we expect to barely see deviations between a LQBH and the corresponding Schwarzschild one for very large $M$'s. We shall come back to these points later.

\subsection{Perturbations of a Scalar Field}

Since in the rest of this paper we shall work only with the background described by the metric (\ref{eq2.8}) and the dimensionless coordinates $\tilde t$ as well as $\tilde r$, without causing any confusion, we shall drop the tildes from the dimensionless coordinates $(\tilde t, \tilde r)$ and the corresponding quantities. 
Then, following \cite{Ramin2021} we consider the scalar perturbations that obey the Klein-Gordon equation \footnote{In this paper, we consider only the relativistic dispersion relation. In principle, higher-order corrections can be also included. But, these high-order terms are of order of ${\cal{O}}(K^n/M_{pl}^{n-2})$ with $n \ge 4$, where $K$ denotes the curvature of the macroscopic BHs, which is quite lower than the Planck scale, $ M_{pl} \gg K$ outsides of such macroscopic BHs. Therefore, for the QNMs of macroscopic BHs, these corrections are negligible. For more details, see, for example, Ref. \cite{Wang2017} and references therein.}
\bqn
\lb{KGeqn}
\frac{1}{\sqrt{-g}} \partial_\mu \left( \sqrt{-g} g^{\mu \nu} \partial_\nu \Phi \right) &=& 0,
\eqn
where $\Phi=\Phi(t, r, \theta, \varphi)$ denotes the scalar field. By decomposing it through the spherical harmonics $Y_{lm}(\theta, \varphi)$ as $\Phi=Y_{lm} \Psi(t, r)/r$, a master equation of $\Psi$ is obtained as
\bqn
\lb{master1}
\frac{\partial^2 }{\partial x^2} \Psi+\left( -\frac{\partial^2}{\partial t^2} +V_{eff}(r) \right)\Psi &=& 0,
\eqn
where the effective potential $V_{\text{eff}}$ is given by 
\bqn
\lb{Veff}
V_{\text{eff}} &\equiv& f \frac{L}{r^2}+\frac{1}{2 r} \frac{d f^2}{dr},
\eqn
with $L\equiv l (l+1)$, and  $x$  the tortoise coordinate defined through 
\bqn
\lb{rstar}
\frac{d r}{d x} &=& f(r).
\eqn
Notice that, since $r$ is  dimensionless, $x$ is actually also  dimensionless. 
Also notice that, when the background has spherically symmetry, the perturbations are independent of the choice of the angular index $m$ \cite{Chandra92}. Thus, without loss of the generality, we shall set $m = 0$.  
For a given function $f(r)$ and choosing a suitable value of $l$, we can solve \eqref{master1} for $\Psi$ (which contains the information of QNMs) for a BH.

\begin{figure}[h]
\includegraphics[width=0.9
\linewidth]{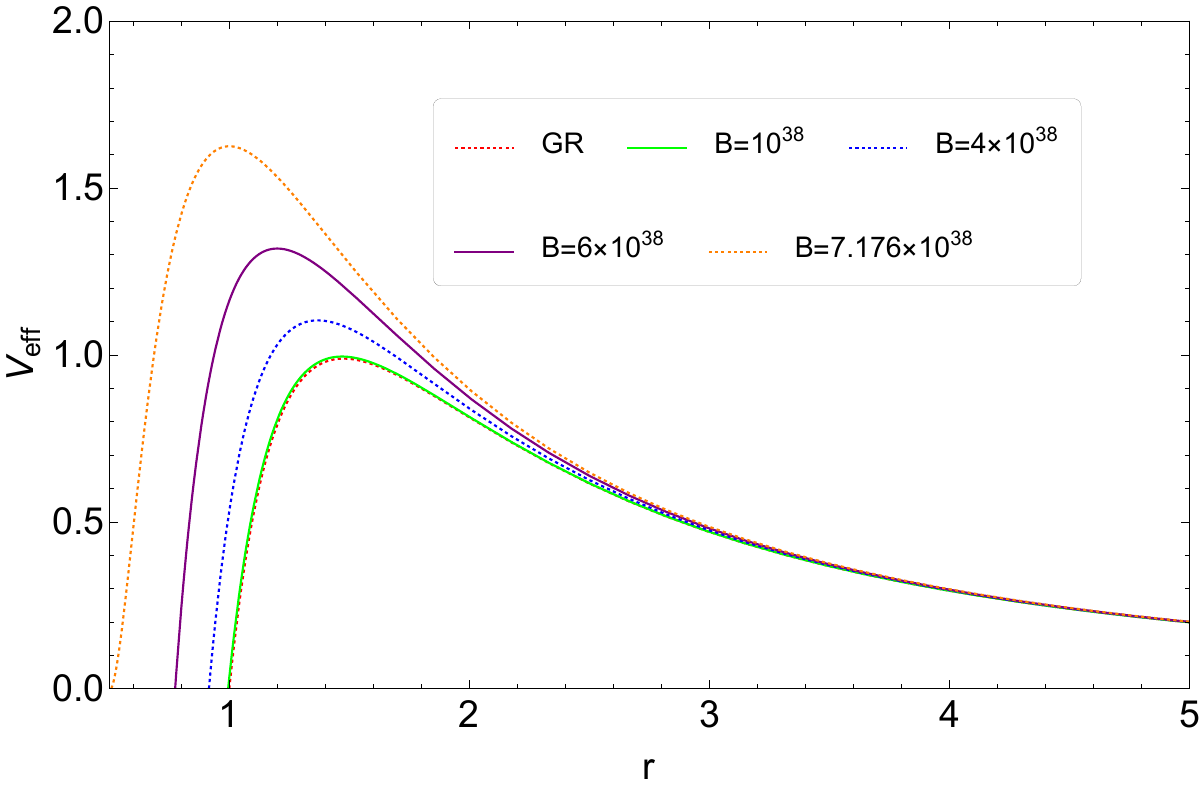} 
\includegraphics[width=0.9
\linewidth]{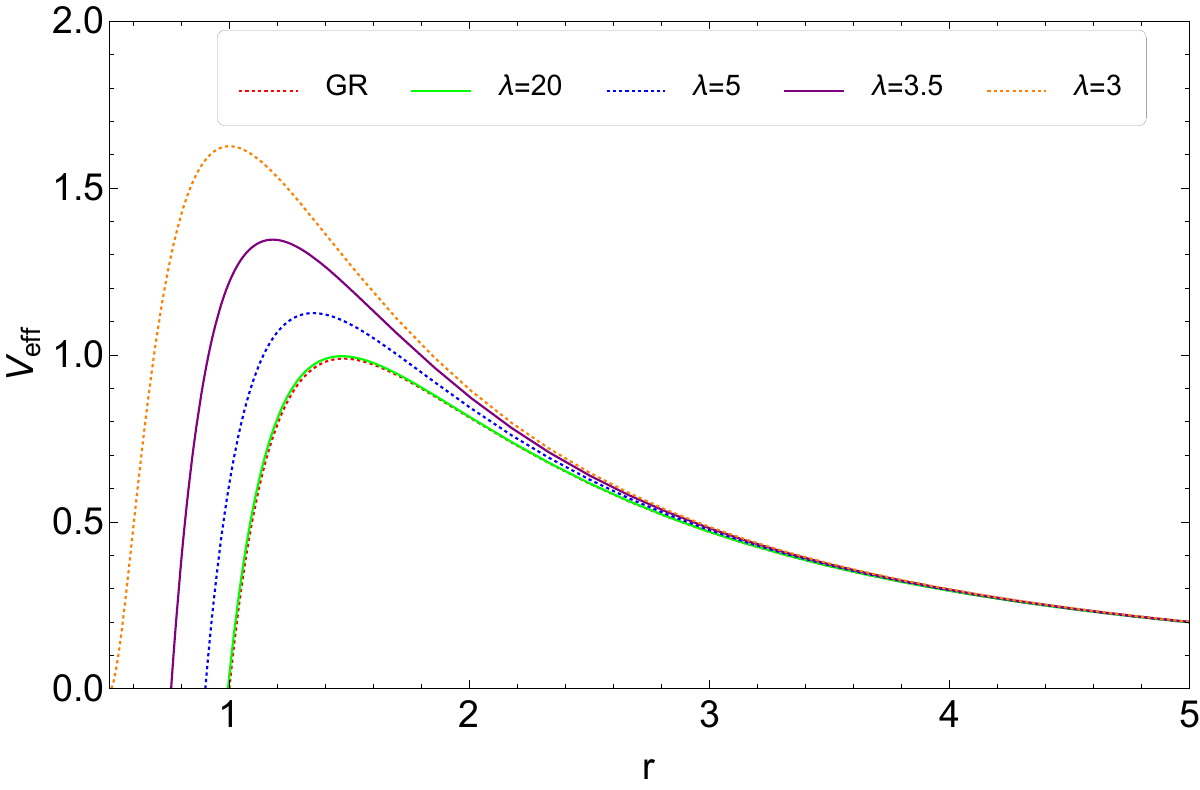} 
\caption{The behavior of $V_{\text{eff}}$ as a function of $r$ (which is in its dimensionless form). Upper panel: For the $\lambda = M/M_{\odot}=3$ and $l=2$ case by varying the parameter $B$. Lower panel: For the $B=7.176 \times 10^{38}$ and $l=2$ case by varying the parameter $\lambda$. The GR case is also exhibited in here as a comparison. 
} 
\label{plotVeff}
\end{figure} 

In fact, the behavior of $V_{\text{eff}}$ is directly related to the results of QNMs. Therefore, to take a glance to the influence of $\{B, \lambda\}$ on the consequence, we first check their influence on $V_{\text{eff}}(r)$.  
The behavior of $V_{\text{eff}}$ is plotted in Fig.\ref{plotVeff} by setting $l=2$. In the upper panel we fix  $\lambda = M/M_{\odot}=3$ and plot out different curves with a varying $B$ factor. In contrast, in the lower panel we fix  $B=B_{max}=7.176 \times 10^{38}$  and plot out different curves with a varying $\lambda \in (3, 100)$. Notice that, the $V_{\text{eff}}$ from GR is added to there as a comparison. 

The upper panel of Fig.\ref{plotVeff} implies that an observable deviation between a LQBH and a Schwarzschild one requires an extremely large $B$, which is consistent with what we observed from Fig.\ref{plot1}.  On the other hand, as expected (again by observing Fig.\ref{plot1}), the lower panel reflects that the deviation on  $V_{\text{eff}}$ from its GR counterpart will soon fade away once $\lambda$ become relatively large (e.g., when it is larger than $20$), even with $B=B_{max}$. 
From the qualitative point of view, Fig.\ref{plotVeff} provides us with a good reference on choosing the parameters from the phase space of $\{B, \lambda\}$ when calculating for QNMs in the next section. Although we are not going to follow Fig.\ref{plotVeff}'s choices precisely, that gives some hints to the choice of the parameters. 
\section{Quasi-Normal Modes of LQBHs}
\renewcommand{\theequation}{3.\arabic{equation}} \setcounter{equation}{0}
\label{secQNM}

First of all, we  try to solve Eq.\eqref{master1} in the time-domain with the finite difference method (FDM) \cite{Chao2023a, Chao2022, RichardB}. 
One advantage of such a method is that, for any chosen $l$, the final result from it can reflect comprehensively the influence of a bunch of different modes (represented by $\{m, n\}$, although $m$ can be always set to zero in the spherical case without loss of the generality, as mentioned earlier. It must not be confused this m with the mass parameter M of the LQBHs.).
To apply the FDM, we first introduce two new variables $\mu \equiv t-x$ and  $\nu \equiv t+x$ [so that  $t=(\nu+\mu)/2$ and $x=(\nu-\mu)/2$]\footnote{ {An important feature of the FDM is that in using it, we need the exact form of $x(r)$, in addition to its derivative with respect to $r$. Therefore, according to the definition \eqref{rstar}, we have to assign $x$ an integral constant to absolutely fix it. In fact, such a constant could be chosen arbitrarily and it's independent of our results for calculating QNMs. Thus, we made a simple choice by letting $x(r=2)=2$, which is consistent with the general choice in GR (See e.g., \cite{Leaver1985}).}}. Therefore, on a $(N+1) \times (N+1)$ lattice (where $N$ is a positive integer that will be chosen properly according to our usage), we perform the calculation of $\Psi(\mu, \nu)$ by using {the recursion formula}
\begin{widetext}
\bqn
\lb{FDM1}
\Psi(\mu+\delta h, \nu+\delta h) &\cong & \Psi(\mu, \nu+\delta h) + \Psi(\mu+\delta h, \nu) - \Psi(\mu, \nu )-\frac{1}{8}\delta h^2 V_{\rm eff}\left(r \right) \left[\Psi(\mu, \nu+\delta h ) + \Psi(\mu+\delta h, \nu)\right],
\eqn
\end{widetext}
where $\delta h$ is the step size (which will be chosen properly in reality by measuring our tolerance of accuracy). The boundary conditions are given by $ \Psi(\mu, \nu=0 ) = 0\;(\mu \ne 0)$ and $ \Psi(\mu=0, \nu ) = \exp[-(\nu-1)^2/2]$. Thus, after $N^2$ iterations, we find all the $\Psi(n \delta h, n \delta h)$'s for $n \in [0, N]\cap \mathbb{Z}$. From that we can calculate $\Psi(t, x=0 )$ by using the relation $\Psi(t, x=0 )=\Psi(\mu, \mu)$.

\begin{figure*}[h]
\includegraphics[width=0.9
\linewidth]{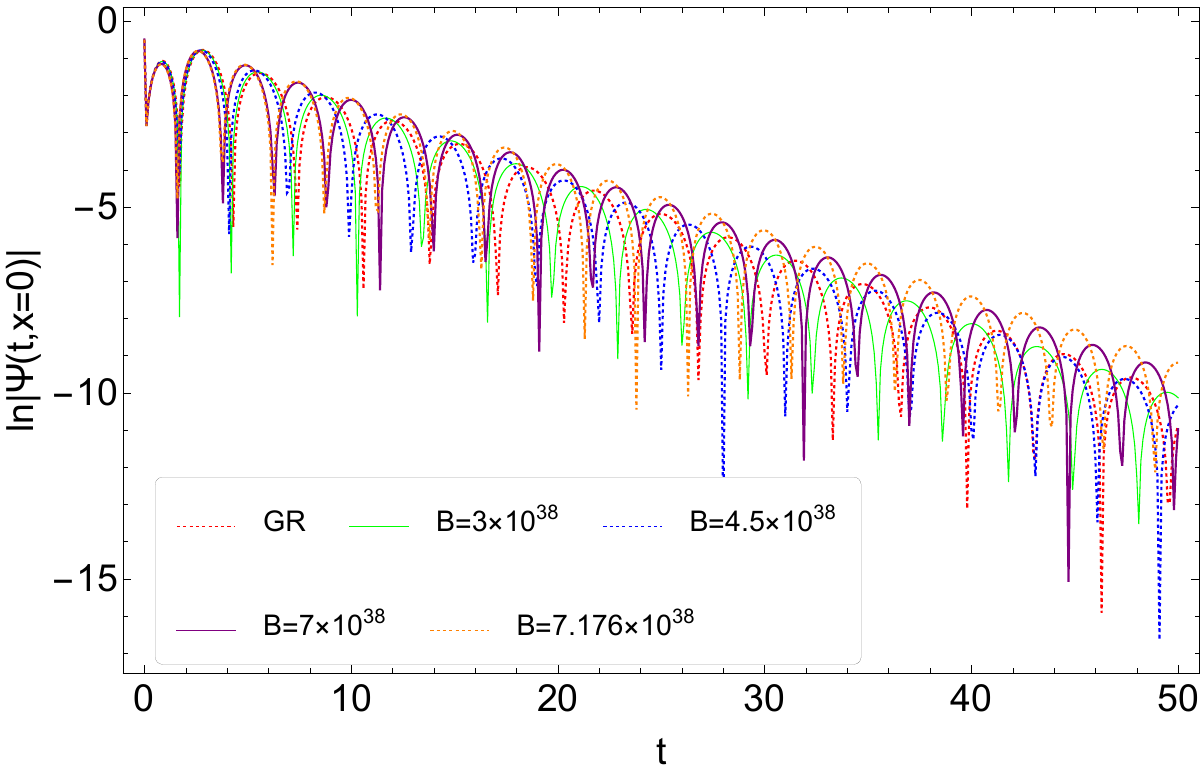} 
\includegraphics[width=0.9
\linewidth]{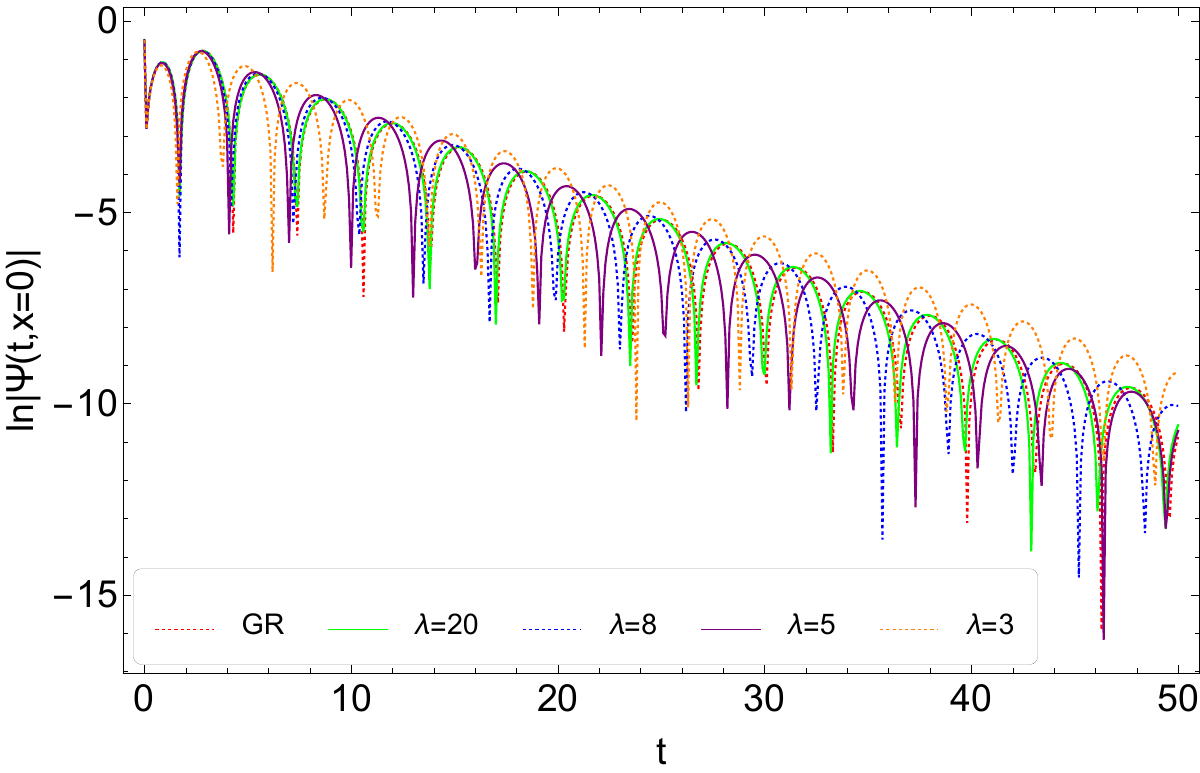} 
\caption{The behavior of $\ln |\Psi(t, x=0)|$ as a function of $t$ (which is in its dimensionless form). Upper panel: For the $\lambda = M/M_{\odot}=3$ and $l=2$ case with the variation of the parameter $B$. Lower panel: For the $B=7.176 \times 10^{38}$ and $l=2$ case with the variation of the  parameter $\lambda$. The GR case is also exhibited in here as a comparison. 
} 
	\label{plot2}
\end{figure*}

Then, the results  are shown in the upper panel of Fig. \ref{plot2}. For the case with $\lambda = M/M_{\odot}=3$ and $l=2$, the behavior of $\ln |\Psi(t, x=0)|$ (which encodes the information of various QNM frequencies) is plotted out as a function of $t$ for different values of $B$ (with these values chosen by referring Fig.\ref{plotVeff}). It should be noted that   the variable $t$ here is dimensionless, which was previously written as ${\tilde t} = t/r_s$.   The GR case is also added as a comparison. 
Just like what we have observed from Fig.\ref{plotVeff}, the result
is not very sensitive to the value of $B$, especially when
it is not large enough, so that the deviation between the
curves of GR (represented by the red dashed line) and
that of the case $B=3\times 10^{38}$ (represented by the green
solid line) is still moderate. 
In contrast, once $B$ reached certain critical values, the difference becomes obvious.  That is why we observe a relatively large deviation between  the curves of the $B=4.5\times 10^{38}$ case (represented by the blue dashed line) and that of the $B=7 \times 10^{38}$ case (represented by the purple solid line). 
On the other hand, by observing the upper panel of  Fig. \ref{plot2} we also notice that the frequency of vibration is getting bigger as the factor $B$ increases.

Similarly, we show the corresponding results when changing $\lambda$ in the lower panel of Fig. \ref{plot2} (with its values chosen by referring Fig.\ref{plotVeff}). By setting $B=7.176 \times 10^{38}$ and $l=2$, the behavior of $\ln |\Psi(t, x=0)|$ is plotted out as a function of $t$. We also add the GR case  in order to make a comparison. As we expect from Fig. \ref{plot1}, we can barely see the deviations between the curve of a LQBH with large masses  (e.g., the one represented by the green solid line) and that of GR (represented by the red dashed line), even if the parameter $B$ is chosen to be $B_{max}$. In contrast, given a relatively small $\lambda$ (e.g., the one represented by the blue dashed line), we can observe a significant deviation.
On the other hand, by observing the lower panel of  Fig. \ref{plot2} we also notice that the frequency of vibrations is getting bigger with the decrease of the factor $\lambda$.

\begin{table*}
\centering
	\caption{Various modes of $\omega$ (in its dimensionless form, for which we dropped the ``tilde'' as mentioned earlier) for LQBHs with different masses that characterized by $\lambda$. The GR limit of $\omega$ for each mode is provided and the percent difference with the GR limit is shown. The factor $B$ is fixed to be $B=7.176 \times 10^{38}$. } 
	\label{table1}   
\begin{tabular}{|c c | c  | c c c | c c c|  c c c|} 
		\hline  
		 &    & GR       &     $\lambda=20$   &    &     &   $\lambda=10$ & &  &   $\lambda=5$ & & 
		\\
		\hline  
		$l$ &   $n$    &  $\omega_{GR}$    &  $\omega$ &   $\Re{(\delta \omega)}$ &    $\Im{(\delta \omega)}$  & $\omega$ &   $\Re{(\delta \omega)}$ &    $\Im{(\delta \omega)}$    & $\omega$ &   $\Re{(\delta \omega)}$ &    $\Im{(\delta \omega)}$ 
		\\
		\hline
		\hline
  		2 &   0   &  $0.96728-0.19353 i$   &  $0.97096  -0.19377   i$ &   $0.4\%$ &    $0.1\%$  & $0.98235 -0.19445  i$ &   $1.6\%$ &    $0.5\%$   & $1.03475  -0.19666   i$ &   $7.0\%$ &    $1.6\%$ 
		\\
  		 2 &   1   &  $0.92769 -0.59125  i$   &  $0.93154  -0.59191    i$ &   $0.4\%$ &    $0.1\%$  & $0.94349 -0.59380   i$ &   $1.7\%$ &    $0.4\%$   & $0.99872  -0.59958   i$ &   $7.6\%$ &    $1.4\%$ 
		\\
      	2 &   2   &  $0.86077  -1.01740   i$   &  $0.86491   -1.01833    i$ &   $0.5\%$ &    $0.1\%$  & $0.87779   -1.02096    i$ &   $2.0\%$ &    $0.4\%$   & $0.93776  -1.02794   i$ &   $8.9\%$ &    $1.0\%$ 
		\\
  		\hline
		3 &   0   &  $1.35073 	-0.19300 i$   &  $1.35585  	-0.19324 i$ &   $0.4\%$ &    $0.1\%$  & $1.37173 	-0.19393 i$ &  $1.6\%$  &    $0.5\%$    & $1.44474  	-0.19618 i$ &  $7.0\%$ &   $1.6\%$ 
		\\
		3 &   1   &  $1.32134 	-0.58458 i$   &  $1.32659  	-0.58526 i$ &   $0.4\%$ &    $0.1\%$  & $1.34287 	-0.58725 i$ &  $1.6\%$ &    $0.4\%$    & $1.41799  	-0.59356 i$ &  $7.3\%$ &    $1.5\%$  
		\\
		3 &   2   &  $1.26718 	-0.99202 i$   &  $1.27267  	-0.99308 i$ &   $0.4\%$ &    $0.1\%$  & $1.28971  	-0.99612 i$ &  $1.8\%$ &    $0.4\%$    & $1.36870  	-1.00519 i$ &  $8.0\%$ &    $1.3\%$  
		\\
  		\hline
    		6 &   0   &  $2.50377 	-0.19261 i$   &   $ 2.51324 - 0.19285 i$ &   $0.4\%$ &     $0.1\%$  & $ 2.54262 	-0.19354 i $ &  $1.6\%$ &    $0.5\%$  & $2.67774  	-0.19582 i$ &  $6.9\%$ &   $1.7\%$ 
		\\
		6 &   1   &  $2.48750 	-0.57947 i$   &  $2.49705 	-0.58018 i $ &   $0.4\%$ &    $0.1\%$  & $2.52665 	-0.58224 i $ &  $1.6\%$ &    $0.5\%$   & $ 2.66293  	-0.58895  i $ &  $7.0\%$ &   $1.6\%$
		\\
		6 &   2   &  $2.45569 	-0.97120 i$   &  $2.46537 	-0.97236 i$ &   $0.4\%$ &    $0.1\%$  & $2.49542  	-0.97572 i $ & $1.6\%$ &    $0.5\%$   & $ 2.63397 	-0.98646 i $ &  $7.3\%$ &   $1.6\%$
		\\
  		\hline
	\end{tabular}
\end{table*}

On the other hand,  Eq.\eqref{master1} can be written in its frequency-domain as \cite{Chao2022}
\bqn
\lb{master2}
\frac{\partial^2 }{\partial x^2} \Psi+\left(\omega^2 +V_{\text{eff}}(r) \right)\Psi &=& 0,
\eqn
 where $\omega$, representing the dimensionless quantity $\tilde\omega$ introduced in Eq.(\ref{eq2.7}),    is the QNM frequency to be calculated. 
To solve for $\omega$ from \eqref{master2}, among various existing methods (see, e.g., \cite{Chandra1975,Kai2017,Ghosh2023}), here we introduce the WKB method \cite{Chao2022}. The QNM frequency can be calculated up to the 6th order by using 
\bqn
\lb{WKB1}
\omega &=& \sqrt{-i \left[\left(n+\frac{1}{2}\right)+\sum_{k=2}^6 \Lambda_k \right] \sqrt{-2 V_0''}+V_0},~~~~
\eqn
where
\bqn
\lb{WKB2}
V_0 \equiv \left. V_{\rm eff} \right|_{r=r_{\text{max}}}, \quad V_0'' \equiv \left. \frac{d V_{\rm eff}}{d r^2} \right|_{r=r_{\text{max}}},
\eqn
\break
with {$V_{\rm eff} (r=r_{\text{max}})$} gives the maximum of {{$V_{\rm eff}$} for $r \in (r_{MH}/r_s,~\infty)$.} The expressions of $\Lambda_k$'s could be found in \cite{Will1985, Iyer1987, Konoplya2003}. Note that $n=0, 1, 2, ...$. 

For the mode $\{l, n\}$ (recall that we we have set $m=0$ so that it is not mentioned), some of the results of $\omega$ are shown in Table \ref{table1}. In this table we fix the factor $B$ as $B=7.176 \times 10^{38}$.  In addition, the GR limit, which is denoted as $\omega_{GR}$, is also added as a comparison. Notice that, to compare with the results given in \cite{Ramin2021}, the modes with $l=2, 6$ are included in Table \ref{table1}. To see clearly the difference between a LQBH and a classical Schwarzschild BH with the same mass, we have introduced the percent difference defined by 
\bq
\lb{eq3.5}
\delta \omega \equiv \frac{\omega-\omega_{GR}}{\omega_{GR}} \times 100 \%,
\eq
with $\Re{(\delta \omega)}$ and $\Im{(\delta \omega)}$ denote the real and imaginary parts
of $\delta \omega$, respectively. 

From Table \ref{table1} we find that the value of $\delta \omega$ is sensitive to the choice of $\lambda$. Just like what we have observed in the lower panel of Fig.\ref{plot2}, a huge $\lambda$ often results in negligible deviations on QNMs between a LQBH and a classical Schwarzschild one.  More importantly, given a sufficiently large $B$ and a reasonable  $\lambda$, the percent difference $\delta \omega$ can well locate in the observational window once the LIGO-Virgo-KAGRA detector network reaches its designed sensitivity \cite{Carullo2018}, whereby some constraints on these LQBHs can be found from GW observations.

In addition, it is worth mentioning here that the corresponding value of $r_{MH}$ for each $\omega$ appearing in Table \ref{table1} effectively reflects its deviation from that of GR (as implied by Fig.\ref{plot1}). Let us take the $l=2$ case as an example. Setting $\lambda=5$ leads to $r_{MH} \approx 0.90003$ while setting $\lambda=20$ leads to $r_{MH} \approx 0.99434$.  Clearly, the latter is quite close to that of GR, for which we have $r_{MH}=1$ [cf., Eq.\ref{ftr}], in comparing to the former. It thus makes sense to us that the former tends to bring us an potentially observable deviation from that of GR as mentioned earlier. Actually, if we enlarge our scope, we shall find that setting $\lambda=3$ will further putting $r_{MH}$  into approximately $0.50800$, which is obviously far away from the GR case.  Although we did not show it explicitly in here, one can easily find out that the corresponding $\omega$'s for $\lambda=3$ are indeed tremendously different from their counterparts of GR. 

\section{Conclusions}
\renewcommand{\theequation}{4.\arabic{equation}} \setcounter{equation}{0}
\label{secconclusion}

In this paper we have investigated QNMs of perturbations
 of a scalar field on the backgrounds of a large
class of LQBHs, which can be formed from gravitational
collapse of realistic matter\cite{Lewandowski2023,Luca2024}. These solutions are
characterized by three parameters, $M$, $\alpha$ and $B$ [cf. Eqs.
\eqref{eqB} and \eqref{alpha}, where $M$ is the mass parameter, and $\alpha$ and $B$ are the two quantum parameters, characterizing
the quantum geometric effects of the LQBHs. To see the effects of these two quantum parameters, it is suggestive to introduce the dimensionless coordinates $({\tilde t}, {\tilde r})$, as defined by Eq.\eqref{eq2.4}, for which the spacetime can be cast in
the form of Eq.\eqref{eq2.8} with

\bqn
f(\tilde{r}) &=& 1 - \frac{1}{\tilde{r}} + \frac{\tilde\alpha}{\tilde{r}^2} \left(\frac{1}{\tilde{r}}+ B\right)^2,
\eqn
where
\bqn
\tilde{\alpha} &\equiv& \frac{\alpha}{4r_s^2} = \sqrt{3} \pi \gamma^3 \left(\frac{\ell_{\text{pl}}}{M}\right)^2  \simeq {\cal{O}}\left(10^{-77}\right),
\eqn
for a solar massive LQBH, $M \simeq 10^{38} l_{pl}$. Therefore, when
$B \simeq 0$, the corresponding LQBHs have negligible effects
to the observations of such BHs. From Eq. \eqref{eqB} it can
be seen that this is indeed the case for $k = 0, +1$. On the other hand, for $k=-1$, we have $B = \sinh^2(R_0)$, where
$R_0$ is the geometric radius of the collapsing dust ball.
Therefore, depending on $R_0$, the parameter $B$ can be
arbitrarily large, so that the leading term ${\tilde \alpha} B^2/{\tilde r}^2$ appearing in $f({\tilde r})$ can have observational effects for macroscopic
LQBHs.

Motivated by the above considerations, we have calculated the QNMs of perturbations of a scalar field in the backgrounds of the above LQBHs. In particular, we have found that they sensitively depend on the ratio
$\lambda = M/M_{\odot}$, when $B$ reaches a critical value, say, $B_c$,
which is of the order of $B_c \simeq M_{\odot}/l_{pl} \simeq 10^{38}$, as shown in Fig.\ref{plotVeff}. Therefore, with $B \simeq B_c$ one can find significant deviations between a LQBH and a Schwarzschild one, for
some values of $M$ (or $\lambda \equiv M/M_{\odot}$) as one can see from
Figs.\ref{plot1} and \ref{plotVeff}.

To be comparable with observations, in this paper we have mainly focused on the cases  $\lambda \in (3, 100)$. For
such an interval of the LQBH masses, there exists an upper bound $B_{max}$ of $B$, only under which LQBHs exist.
We have found that this maximal value is $B_{max} \approx 7.176 \times 10^{38}$. On the other hand, we have also found that, working with the dimensionless coordinates $({\tilde t}, {\tilde r})$ can significantly simplify the calculations of the QNMs of the LQBHs. With this in mind, we have first studied the
perturbations of the time-domain master equation \eqref{master1}.
Using the FDM method [cf., \eqref{FDM1}], we have solved the
equation and plotted the behavior of $\ln |\Psi(t, x=0)|$ as
a function of $\tilde t$ in Fig.\ref{plot2} (in which simply written as $t$) for the dominant mode $l = 2$. In the upper panel of this figure, we have fixed $\lambda = 3$,
while varying the parameter $B$. On the other hand, in the lower panel, we have fixed $B = 7.176 \times 10^{38}$ but now varying $\lambda$. Then, we have shown that only in the neighborhood of $(B, \lambda) \approx (B_c, 3)$ in the phase space, can the deviations of the QNMs between LQBHs and the corresponding classical BH become significant.

To find out more details, we have further calculated explicitly the values of $\omega$ by solving the master equation in the frequency-domain [cf., \eqref{master2}]. The corresponding results are exhibited in Table \ref{table1}. By examining the percent difference between the classical and LQG BHs, defined by \eqref{eq3.5}, we have shown that the percent difference as large as $2\%$ (or larger) can be obtained for various choices of  $\{B, \lambda\}$. These differences could be well within the detectability of the forthcoming third-generation detectors
of GWs \cite{Bustillo2021,Shi2024}, whereby one can either rule out some regions of the phase space of the LQBHs or confirm them.

Finally, we note that our current work can be extended to several different directions. For instance, one may study the QNMs of other LQBHs considered in \cite{OG2024}. On the other hand, one may study the QNMs of metric perturbations of LQBHs recently developed in \cite{Mena2024}.

\section*{Acknowledgments}

We thank Profs. J. Lewandowski and P. Singh for valuable
 discussions on loop quantum black holes. We also thank Dr. Kai Lin for valuable discussions on the calculations  of QNMs. C.Z. is
supported in part by the National Natural Science Foundation
 of China under Grant No. 12205254, and the Science
 Foundation of China University of Petroleum, Beijing
 under Grant No. 2462024BJRC005. A.W. is partially
 supported by the US National Science Foundation
(NSF) through the grant: PHY-2308845.




\begin{thebibliography}{nbound}




\bibitem{test_GR1}
E. Berti, E. Barausse, V. Cardoso, L. Gualtieri, P. Pani, {\rm et. al.},  Testing General Relativity with Present and Future Astrophysical Observations, Class. Quantum Grav. {\bf 32}, 243001 (2015).

\bibitem{test_GR2}
L. Barack, V. Cardoso, S. Nissanke, T. P. Sotiriou, {\em et. al.}, Black Holes, Gravitational Waves and Fundamental Physics: A Roadmap, Class. Quantum Grav. {\bf 36}, 143001 (2019).

\bibitem{test_GR3}
The LIGO Scientific Collaboration, the Virgo Collaboration, the KAGRA Collaboration,, Tests of General Relativity with GWTC-3, arXiv: 2112.06861 [astro-ph].


\bibitem{Xiang2019}  X. Zhao, C. Zhang, K. Lin, T. Liu, R. Niu, B. Wang, S.-J.  Zhang, X. Zhang, W. Zhao, T. Zhu, A. Wang, Gravitational waveforms and radiation powers of the triple  system PSR J0337+1715 in modified theories of gravity, Phys. Rev. D{\bf 100}, 083012 (2019).

\bibitem{Chao2020}  C. Zhang, X. Zhao,  A. Wang, B. Wang, K. Yagi, N. Yunes, W. Zhao and T. Zhu, Gravitational waves from the quasicircular inspiral of compact binaries in Einstein-aether theory, Phys. Rev. D{\bf 101}, 044002 (2020).	

\bibitem{Chao2020b} C. Zhang, X. Zhao, K. Lin, S.-J. Zhang, W. Zhao and A.-Z. Wang, Spherically symmetric static black holes in Einstein-aether theory, Phys. Rev. D{\bf 102}, 064043 (2020).

\bibitem{Chao2023a}
C. Zhang, A. Wang and T. Zhu, Odd-parity perturbations of the
wormhole-like geometries and quasi-normal modes in Einstein-Æther theory,  JCAP {\bf 05} (2023) 059.

\bibitem{Zack2020} Zack Carson and Kent Yagi, Probing Einstein-dilaton Gauss-Bonnet gravity with the inspiral and ringdown of gravitational waves, Phys. Rev. D{\bf 101}, 104030  (2020).

\bibitem{Schwarzschild1916}
K. Schwarzschild, Sitzungsber. Preuss. Akad. Wiss. Berlin (Math. Phys. ) {\bf 1916} 189 (1916).

\bibitem{Ref1} B.P. Abbott, {\it et al.,} [LIGO/Virgo Scientific Collaborations], Observation of Gravitational Waves from a Binary Black Hole Merger, Phys. Rev. Lett. {\bf 116},  061102 (2016).


 


\bibitem{GWs}  B.P. Abbott, {\it et al.,} [LIGO/Virgo Collaborations], GWTC-1: A Gravitational-Wave Transient Catalog of Compact Binary Mergers Observed by LIGO and Virgo during the First and Second Observing Runs, Phys.  Rev. X{\bf 9},  031040 (2019).

\bibitem{GWs19a}  B.P. Abbott, {\it et al.,} [LIGO/Virgo Collaborations], Open data from the first and second observing runs of Advanced LIGO and Advanced Virgo, SoftwareX, Volume 13, 100658 (2021).

\bibitem{GWs19b} B.P. Abbott, {\it et al.,} [LIGO/Virgo Collaborations], GW190425: Observation of a Compact Binary Coalescence with Total Mass $\sim 3.4M_{\bigodot}$, ApJL {\bf 892} L3 (2020);
\url{https://www.ligo.caltech.edu}.

\bibitem{GWsO3b} B.P. Abbott, {\it et al.,} [LIGO/Virgo/KAGRA Collaborations], GWTC-3: Compact Binary Coalescences Observed by LIGO and Virgo During the Second Part of the Third Observing Run,
arXiv:2111.03606v1 [gr-qc].

\bibitem{Moore2015}  C. J. Moore, R. H. Cole and C. P. L. Berry,  Gravitational-wave sensitivity curves, Class. Quantum. Grav. {\bf 32}, 015014 (2015).


\bibitem{Gong:2021gvw}
Y.~Gong, J.~Luo and B.~Wang,
``Concepts and status of Chinese space gravitational wave detection projects,''
Nature Astron. \textbf{5}, no.9, 881-889 (2021).

\bibitem{CE} 
\url{https://cosmicexplorer.org}.

\bibitem{ET} ET Steering Committee Editorial Team, ET design report update 2020, ET-0007A- 20 (2020); \url{https://www.et-gw.eu/}.

\bibitem{LISA} 
\url{https://www.lisamission.org}.

\bibitem{Liu2020} S. Liu, Y. Hu, {\it et al.}, Science with the TianQin observatory: Preliminary results on stellar-mass binary black holes, Phys. Rev. D{\bf 101}, 103027 (2020).

\bibitem{Shi2019} C.-F. Shi,  {\it et al.}, Science with the TianQin observatory: Preliminary results on testing the no-hair theorem with ringdown signals, Phys. Rev. D{\bf 100}, 044036 (2019).

\bibitem{Taiji2} W.-H. Ruan, Z.-K. Guo, R.-G. Cai, Y.-Z. Zhang, Taiji Program: Gravitational-Wave Sources, Int. J. Mod. Phys. A {\bf 35}, No. 17, 2050075 (2020).

\bibitem{DECIGO} S. Kawamura, {\it et al.,} Current status of space gravitational wave antenna DECIGO and B-DECIGO, arXiv:2006.13545.

\bibitem{Berti2009} E. Berti, V. Cardoso and A. O. Starinets,  Quasinormal modes of black holes and black branes, Class. Quantum. Grav. {\bf 26}, 163001 (2009).

\bibitem{Berti18}  E. Berti, K. Yagi, H. Yang, N. Yunes,  Extreme gravity tests with gravitational waves from compact binary coalescences: (II) ringdown, Gen. Relativ. Grav. {\bf 50},  49 (2018).

\bibitem{Gong2023}
C. Zhang, Y. Gong, D. Liang and B. Wang, Gravitational waves from eccentric extreme mass-ratio inspirals as probes of scalar fields,  JCAP {\bf 06} (2023) 054.

\bibitem{Tu2023} Ze-Yi Tu, Tao Zhu and Anzhong Wang, Periodic orbits and their gravitational wave radiations in a polymer black hole in loop quantum gravity, Phys. Rev. D{\bf 108}, 024035  (2023).

\bibitem{Chao2022}
C. Zhang, T. Zhu, X. Fang and A. Wang, Imprints of dark matter on gravitational ringing of supermassive black holes, Physics of the Dark Universe {\bf 37}, 101078 (2022).

\bibitem{Shen2024}
Z. Shen, A. Wang, Y.-G. Gong, S.-Y. Yin, Analytical
 models of supermassive black holes in galaxies surrounded
 by dark matter halos, Phys. Lett. B{\bf 855} (2024) 138797.
 



\bibitem{Cheung2023}
M.H.Y. Cheung, {\it et al.,} Nonlinear effects in black hole
ringdown, Phys. Rev. Lett. {\bf 130}, No. 8 (2023) 081401.

\bibitem{Mitman2023}
K. Mitman, {\it et al.,} Nonlinearities in Black Hole
Ringdowns, Phys. Rev. Lett. {\bf 130}, 081402 (2023).

\bibitem{Regge1957}
T. Regge and J.A. Wheeler, Stability of a Schwarzschild singularity,
 Phys. Rev. {\bf 108} (1957) 1063.
 
\bibitem{Teukolsky1973}  
S. A. Teukolsky, Perturbations of a Rotating Black Hole. I. Fundamental Equations for Gravitational, Electromagnetic, and Neutrino-Field Perturbations, Astrophys. J. {\bf 185}, 635
(1973).

\bibitem{Will1985} 
B. F. Schutz and C. M. Will, BLACK HOLE NORMAL MODES: A SEMIANALYTIC APPROACH, Astrophys. J.  {\bf 291},  L33-L36 (1985).

\bibitem{Leaver1985}
E. W. Leaver, An analytic representation for the quasi-normal modes of Kerr black holes, Proc. R. Soc. Lond. A. {\bf 402}, 285-298 (1985).

\bibitem{Iyer1987}
S. Iyer and C. M. Will, Black-hole normal modes: A WKB approach. I. Foundations and application of a higher-order WKB analysis of potential-barrier scattering, Phys. Rev. D{\bf 35}, 12 (1987); S. Iyer, Black-hole normal modes: A WKB approach. II. Schwarzschild black holes, Phys. Rev. D{\bf 35}, 3632  (1987).

\bibitem{Chandra92} 
S. Chandrasekhar, the mathematical theory of black holes, Oxford classic texts in the physical sciences (Oxford Press, Oxford, 1992).

\bibitem{Kokkotas1999} 
K.D. Kokkotas, B.G. Schmidt, Quasi-normal modes of
stars and black holes. Living Reviews in Relativity {\bf 2}
(1999) 1.

\bibitem{Konoplya2011} 
R. Konoplya, A. Zhidenko, Quasinormal modes of black
holes: From astrophysics to string theory. Reviews of
Modern Physics {\bf 83} (2011) 793.




\bibitem{Carullo2018} 
Gregorio Carullo, {\it et al.,} Empirical tests of the black hole no-hair conjecture using gravitational-wave observations, Phys. Rev. D{\bf 98}, 104020 (2018).

\bibitem{Isi2019} 
M. Isi, M. Giesler, W. M. Farr, M. A. Scheel and S. A.
Teukolsky, Testing the no-hair theorem with GW150914,
Phys. Rev. Lett. {\bf 123}, 111102 (2019).

\bibitem{Bustillo2021} 
J.C. Bustillo, P.D. Lasky, and E. Thrane, Black-hole
spectroscopy, the no-hair theorem, and GW150914:
Kerr versus Occam, Phys. Rev. D{\bf 103}, 024041 (2021).

\bibitem{Finch2022} 
E. Finch and C. J. Moore, Searching for a ringdown
overtone in GW150914, Phys. Rev. D{\bf 106}, 043005 (2022).

\bibitem{Ma2023} 
S. Ma, L. Sun, and Y. Chen, Using rational filters to uncover the first ringdown overtone in GW150914, Phys. Rev. D{\bf 107}, 084010 (2023).

\bibitem{Isi2023}
M. Isi and W. M. Farr, Comment on “Analysis of Ringdown
 Overtones in GW150914”, Phys. Rev. Lett. {\bf 131},
169001 (2023).

\bibitem{Carullo2023} 
G. Carullo, R. Cotesta, E. Berti, and V. Cardoso, Reply
 to Comment on “Analysis of Ringdown Overtones
in GW150914”, Phys. Rev. Lett. {\bf 131}, 169002 (2023).

\bibitem{Capano2023} 
C. D. Capano, M. Cabero, J. Westerweck, J. Abedi, S.
Kastha, A. H. Nitz, Y.-F. Wang, A. B. Nielsen, and B.
Krishnan, Multimode Quasinormal Spectrum from a Perturbed
 Black Hole, Phys. Rev. Lett. {\bf 131}, 221402 (2023)

\bibitem{Shi2024} 
C.-F. Shi, Q.-F. Zhang, and J.-W. Mei, On the detectability
 and resolvability of quasi-normal modes
with space-based gravitational wave detectors,
arXiv:2407.13110 [gr-qc].



\bibitem{Lewandowski2023} J. Lewandowski, Y. Ma, J. Yang and C. Zhang, Quantum Oppenheimer-Snyder and Swiss Cheese Models, Phys.
Rev. Lett. {\bf 130} (2023) 101501.


\bibitem{Luca2024} Luca Cafaro and Jerzy Lewandowski, Status of Birkhoff's theorem in polymerized semiclassical regime of Loop Quantum Gravity,  arXiv:2403.01910.

\bibitem{Gambini2023} R. Gambini, J. Olmedo, J. Pullin, Quantum geometry
and black holes, in C. Bambi, L. Modesto, I. Shapiro,
(eds) Handbook of Quantum Gravity (Springer, Singapore
 (2023)).
 
\bibitem{Ashtekar2023} A. Ashtekar, J. Olmedo, P. Singh, Regular black holes from Loop Quantum Gravity, in C. Bambi, L. Modesto, I.
Shapiro, (eds) Handbook of Quantum Gravity (Springer,
Singapore (2023)).

\bibitem{OG2024}  G. Ongole, P. Singh, A. Wang, Revisiting quantum black
holes from effective loop quantum gravity, Phys. Rev.
D{\bf 109}, 026015 (2024);
W.-C. Gan, G. Ongole, E. Alesci, Y. An, F.-W. Shu, A. Wang, Understanding
 quantum black holes from quantum reduced
loop gravity, Phys. Rev. D{\bf 106}, 126013 (2022);
W.-C. Gan, N.O. Santos, F.-W. Shu, A. Wang, Properties of the
spherically symmetric polymer black holes, Phys. Rev.
D{\bf 102}, 124030 (2020).

\bibitem{GZO2024}  W.-C. Gan, X.-M. Kuang, Z.-H. Yang, Y.-G. Gong, A.
Wang, B. Wang, Nonexistence of quantum black and
white hole horizons in an improved dynamic approach,
Science China: Phys., Mech. $\&$ Astron. {\bf 67}, No. 8, 280411
(2024);
H.-C. Zhang, W.-C. Gan, Y.-G. Gong, A. Wang, On the improved dynamics
approach in loop quantum black holes, Commun. Theor.
Phys. {\bf 76}, 035401 (2024); 
G. Ongole, H.-C. Zhang, T. Zhu, A. Wang and B. Wang, Dirac Observables in the 4-Dimensional Phase Space of Ashtekar’s Variables and
Spherically Symmetric Loop Quantum Black Holes, Universe
{\bf 2022}, 8, 543 (2022).

\bibitem{Yan2023}
Jian-Ming Yan, {\it et al.}, Observational tests of a quantum extension of Schwarzschild spacetime in loop quantum gravity with stars in the Galactic Center, Phys. Rev. D {\bf 107}, 084043
(2023).

\bibitem{Yan2022} J.-M. Yan, Q. Wu, C. Liu, T. Zhu, A. Wang, Constraints
on self-dual black hole in loop quantum gravity with S0-2 star in the galactic center, JCAP, {\bf 09} (2022) 008.

\bibitem{Liu2021} Y.-C. Liu, J.-X. Feng, F.-W. Shu, A. Wang, Extended
geometry of Gambini-Olmedo-Pullin polymer black hole
and its quasinormal spectrum, Phys. Rev. D{\bf 104} (2021)
106001.

\bibitem{Ramin2021} Ramin G. Daghigh, Michael D. Green and Gabor Kunstatter, Scalar perturbations and stability of a loop quantum corrected Kruskal black hole, Phys. Rev. D{\bf 103}, 084031  (2021).

\bibitem{Arbey2021} A. Arbey, J. Aufffnger, M. Geiller, E.R. Livine, F. Sartini, Hawking radiation by spherically-symmetric static black
holes for all spins: I – Teukolsky equations and potentials,
Phys. Rev. D{\bf 103}, 104010 (2021).

\bibitem{Livine2024}
 E.R. Livine, C. Montagnon, N. Oshita, H. Roussille,
Scalar Quasi-Normal Modes of a Loop Quantum Black
Hole, [arXiv:2405.12671 [gr-qc]].

\bibitem{Xia2024} Z.-W.Xia, H. Yang, Y.-G. Miao, Scalar ffelds around a rotating loop quantum gravity black hole: Waveform, quasi-normal modes and superradiance, Class. Quantum Grav. {\bf 41} 165010 (2024).

\bibitem{Fu2023}  G. Fu, D. Zhang, P. Liu, .-M. Kuang, J.-P. Wu, Peculiar
properties in quasi-normal spectra from loop quantum
gravity effect, arXiv:2301.08421 [gr-ac].

\bibitem{Shao2024}  C.-Y. Shao, C. Zhang, W. Zhang, and C.-G. Shao, Scalar
ffelds around a loop quantum gravity black hole in de
Sitter spacetime: Quasinormal modes, late-time tails
and strong cosmic censorship, Phys. Rev. D{\bf 109}, 064012
(2024).

\bibitem{Yang2023} J. Yang, C. Zhang, Y. Ma, Shadow and stability of
quantum-corrected black holes, Eur. Phys. J. C{\bf 83} (2023)
619.
 
 
 \bibitem{Okounkova2020} M. Okounkova, L.C. Stein, J. Moxon, M.A. Scheel, S.A.Teukolsky, Numerical relativity simulation of GW150914
beyond general relativity, Phys. Rev. D{\bf 101}, 104016
(2020).

\bibitem{LiuC2020} C. Liu, T. Zhu, Q. Wu, {\it et al.}, Shadow and Quasinormal Modes of a Rotating Loop Quantum Black Hole, Phys. Rev. D{\bf 101}, 084001 (2020); Phys.Rev. D{\bf 103} (2021) 089902 (erratum).


\bibitem{Cao:2024oud} L.-M. Cao, J.-N. Chen, L.-B. Wu, L. Xie, Y.-S. Yu,  
    The pseudospectrum and spectrum (in)stability of quantum corrected Schwarzschild black hole,
    arXiv:2401.09907. 



\bibitem{Mena2024} 
G.A. Mena Marug{\'a}n, A. M{\'i}nguez-S{\'a}nchez, Axial perturbations
 in Kantowski-Sachs spacetimes and hybrid quantum
 cosmology, Phys.Rev.D{\bf 109}, No. 10 (2024), 106009.

\bibitem{Stashko2024} Oleksandr Stashko,  Quasinormal modes and gray-body factors of regular black holes in asymptotically safe gravity, arXiv:2407.07892.

\bibitem{Chiou2008} Dah-Wei Chiou, Phenomenological loop quantum geometry of the Schwarzschild black hole, Phys. Rev. D{\bf 78}, 064040  (2008).

\bibitem{Meissner2004}  Krzysztof A Meissner,  Black-hole entropy in loop quantum gravity, Class. Quantum. Grav. {\bf 21},  5245–5251 (2004).

\bibitem{LVK2024} The LVK Scientific Collaboration,  Observation of Gravitational Waves from the Coalescence of a 2.5-4.5 $M_{\odot}$ Compact Object and a Neutron Star, arXiv:2404.04248 (2024).

\bibitem{RichardB}
R. Haberman, APPLIED PARTIAL DIFFEREN-
TIAL EQUATIONS: with Fourier Series and Boundary
Value Problems (5th ed.) (Pearson Education, Inc., One
Lake Street, New Jersey 07458, USA, 2013).

\bibitem{Chandra1975} S. Chandrasekhar, F. R. S., and S. Detweiler, The quasi-normal modes of the Schwarzschild black hole, Proc. R. Soc. Lond. A. {\bf 344}, 411-452 (1975).

\bibitem{Kai2017}
K.  Lin and W.-L. Qian, A matrix method for quasinormal modes: Schwarzschild black holes
in asymptotically flat and (anti-) de Sitter spacetimes, Class. Quantum Grav. {\bf 34}, 095004 (2017).

\bibitem{Ghosh2023} R. Ghosh, N. Franchini, {\it et al.}, Quasinormal modes of nonseparable perturbation equations: The scalar non-Kerr case, Phys. Rev. D{\bf 108}, 024038 (2023).

\bibitem{Konoplya2003} R. A. Konoplya, Quasinormal behavior of the D-dimensional Schwarzschild black hole and the higher order WKB approach, Phys. Rev. D{\bf 68}, 024018 (2003).

 

\bibitem{Wang2017} A. Wang, Ho\v{r}ava gravity at a Lifshitz point: a progress report, Inter. J. Mod. Phys. D{\bf 26} (2017) 1730014. 
 






\end{thebibliography}
\end{document}